\documentstyle[12pt,psfig]{article}

\newcommand{\li}{\limits}
\newcommand{\lra}{\longrightarrow}
\newcommand{\nn}{\nonumber\\}

\newcommand{\tr}{\mbox{tr }}
\newcommand{\pderiv}[1]{\frac{\partial}{\partial #1}}

\newcommand{\sect}[1]{\mbox{$\Gamma(#1)$}}
\newcommand{\map}[5]{\mbox{$\begin{array}{cccc} #1:& #2 & \lra & #3 \\
                                               & #4 & \longmapsto & #5 
                           \end{array}$}}

\newcommand{\Z}{Z\!\!\!Z}
\newcommand{\Ibb}[1]{ \mbox{\rm I\ifmmode\mkern
            -3.6mu\else\kern -.2em\fi#1}}
\newcommand{\ibb}[1]{\leavevmode\hbox{\kern.3em\vrule
     height 1.2ex depth -.3ex width .2pt\kern-.3em\mbox{\rm#1}}}
\newcommand{\C}{{\ibb C}}
\newcommand{\R}{{\Ibb R}}

\newcommand{\dirac}{/\mkern -10mu \partial}

\newcommand{\be}{\begin{equation}}
\newcommand{\ee}{\end{equation}}
\newcommand{\bea}{\begin{eqnarray}}
\newcommand{\eea}{\end{eqnarray}}

\begin{document}
\renewcommand{\thefootnote}{\fnsymbol{footnote}}
\noindent
{\Large\bf Topological Investigation of the Fractionally \\ Quantized Hall Conductivity}\\[1cm]
{\large T. Asselmeyer\footnote[1]{email:torsten@summa.physik.hu-berlin.de}$\;$ and R. Keiper} \\[0.5cm]
{\large Institute of Physics, Humboldt University Berlin \footnote[2]{Invalidenstr. 110, D - 10115 Berlin}, Germany}\\[2cm]
\renewcommand{\thefootnote}{\arabic{footnote}}

\noindent
{\bf Abstract.} Using the fiber bundle concept developed in geometry and topology, the fractionally quantized Hall conductivity is discussed in the relevant many--particle configuration space. Electron-magnetic field and electron-electron interactions under FQHE conditions are treated as functional connections over the torus, the torus being the underlying two-dimensional manifold. Relations to the $(2+1)$--dimensional Chern--Simons theory are indicated. The conductivity being a topological invariant is given as $\frac{e^2}{h}$ times a linking number which is the quotient of the winding numbers of the self-consistent field and the magnetic field, respectively. Odd denominators are explained by the two spin structures which have been considered for the FQHE correlated electron system. \\[0.5cm]
submitted to Annalen der Physik \\[0.3cm]

\setcounter{footnote}{0}
\section{Introduction}

The quantum Hall effect (QHE) is a macroscopic quantum phenomenon that appears in quasi-two dimensional electron systems in a strong magnetic field at very low temperatures \cite{MacDon:89,Sto:92,Haj:94}. This phenomenon shows a certain universality, which means that in definite limits the QHE is generally independent of the material and geometry of the sample. In experiment, high charge mobility and disorder (the random background potential) must be optimized, in order to get well structured spectra of the non-diagonal and diagonal resistivities $\rho_{xy}$ and $\rho_{xx}$, respectively. Depending on the ratio between the 2d-electron concentration ${\cal N}_s$ and magnetic flux density $B$, the relevant Landau level filling factor $\nu=\frac{h}{e}\frac{{\cal N}_s}{B}$ is observed in plateaus which form quantized series $\nu=\frac{p}{q}$ where $p$ and $q$ are integers and $q$ is an odd number.

The integer quantum Hall effect (IQHE) can be explained within the one-electron picture. In this picture, Landau energy levels, which are broadened by disorder, are successively filled with electrons. An understanding of the fractional quantum Hall effect (FQHE), on the other hand, must take electron correlation into account. Because there is no small parameter in the considered problem one cannot use any standard methods of perturbation theory. Following the concept of quantum interference of charges or magnetic fluxes (vortices) on mesoscopic length scales, one looks for a suitable mathematical description of the strongly correlated system.

A very fruitful approach to FQHE theory has already been given in 1983 by Laughlin who introduced a many-particle trial wave function which describes electron correlation by including only degenerate states of the lowest, spin-polarized Landau level \cite{Lau:83}. In 1989 Jain generalized the Laughlin incompressible quantum liquid state to a construction including higher Landau level states, giving rise to the so-called composite fermion picture of the FQHE \cite{Jai:89}. However, many aspects of the formation and interaction of these new quasi-particles have remained unclear up to now. In the present paper we are going to discuss the dynamics of the relevant correlated system by using topological methods. In doing this, we hope to illustrate some fundamental features of the behaviour of interacting 2d-electrons in a strong magnetic field.

The paper is organized as follows. In section 2 we discuss the relation between the degenerate one-electron states in a magnetic field and the flux quantization which is generally a topological property of the system. Here the electron spin has been considered as done by Semenoff and Sodano in 1986 \cite{SeSo:86}. Section 3 deals with the appropriate description of the dynamic electron-electron interaction as a functional connection in a bundle of Hilbert spaces. Here we confirm new results of Flohr and Varnhagen concerning ${\cal W}_{1+\infty}$ symmetry of the FQHE system \cite{FlVa:94}. In section 4 we extend the topological treatment of electron correlation to a second spin structure. A corresponding discussion recently given in a paper of Asselmeyer and He\ss$\;$may explain the odd numbered denominators of fractional filling factors in more detail \cite{AsHe:95}. In section 5 the fractional Hall conductivity is calculated as the current-current correlator by means of (i) the adiabatic curvature and (ii) a generating functional in (2+1)-dimensional field theory. Doing this we hope to give a certain foundation for the composite fermion picture mentioned above. A summary and related conclusions are given in section 6.

\section{Two-dimensional electrons and flux quanta}

As it is well known, the stationary Schr\"odinger equation for a 2d-electron in a magnetic field $\vec{B}=(0,0,B)$ can be exactly solved. In the case of an asymmetric gauge for the vector potential $\vec{A}=(-By,0,0)$ and periodic boundary conditions in the $x$--direction one finds the following Landau states
\be
\Psi_{N\, k_x}(x,y) = C_N e^{-\frac{1}{2}\left(\frac{y-y_0}{l_0}\right)^2} H_N \left(\frac{y-y_0}{l_0}\right)  e^{ik_x x}
\ee
where $C_N$ is a renormalization factor, $H_N$ denotes the Hermitian polynomials, $l_0=\left(\frac{\hbar}{eB}\right)^{\frac{1}{2}}$ is the magnetic length, and $y_0=l_0^2k_x$ is the $y$-coordinate of the classical centre of cyclotron motion, which does not commute with the corresponding coordinate $x_0$. The eigenenergies are those of the harmonic oscillator, depend only on the Landau quantum number $N$, and are degenerate with respect to the quasi momentum $\hbar k_x$ . Because no periodic gauge potential exists, one has to couple translation with gauge transformation in order to fulfill the periodic boundary condition also in y--direction, i.e.
\be
\Psi_{N\, k_x}(x,y) \lra \Psi'_{N\, k_x}(x,y)=e^{i\frac{e}{\hbar}ByL_x} \Psi_{N\, k_x}(x,y) \label{gauge1}
\ee
with a $y$--dependent phase factor. The periodicity condition is now
\be
\frac{e}{\hbar}BL_xL_y=2\pi \gamma \qquad \gamma = 1,2,\ldots\;\;\in\Z\label{quant1}
\ee
and $\gamma = \frac{L_xL_y}{2\pi l_0^2}$ denotes the degree of degeneracy for each Landau level with respect to the plane $L_xL_y$. That means that every Landau level accommodates $\gamma$ electron states. For a total electron number $N_e$, the filling factor is $\nu=\frac{N_e}{\gamma}$, the number of occupied Landau levels. 

From a more mathematical point of view, two remarks concerning periodic boundary conditions should be of interest. First, equations (\ref{gauge1}) and (\ref{quant1}) are analogous to the problem of Hofstadter \cite{Ho:76}, who considered the behaviour of magnetic field dependent wave functions and eigenenergies on a strongly periodic lattice of atomic spacing. Our case, on the contrary, is characterized by the larger periodicity lengths $L_x, L_y$. Second, fulfilling periodic boundary conditions for a rectangular plane $L_xL_y$ is equivalent to changing the underlying manifold from $\R^2$ to the torus $T^2$, which is twice connected.

In the case of a symmetric gauge $\vec{A}=\frac{1}{2}[\vec{B}\times \vec{r}]$ and an exponential decay of the wave function in radial direction, one has electron states of the form
\be
\Psi_{nm}(\vec{r})=\frac{1}{l_0} \tilde{R}_{nm}(r) \frac{1}{\sqrt{2\pi}} e^{im\phi}\label{solution1} \qquad .
\ee
In the dimensionless variable $\xi =\frac{1}{2}\left(\frac{r}{l_0}\right)^2$ the radial functions are given by
\be
\tilde{R}_{nm}(\xi) = \left[ \frac{n!}{(n+|m|)!}\right]^{\frac{1}{2}} e^{-\frac{\xi}{2}} \xi^{\frac{|m|}{2}} L_n^{|m|}(\xi) \qquad ,
\ee
 $L_n^{|m|}(\xi)$ being the Laguerre polynomials. The Landau levels depend on the main quantum number $N=n+\frac{1}{2}(|m|+m)\; ,\, n=0,1,2,\ldots\;$; where $\, m=-\infty ,\ldots ,0,\ldots ,N$ now expresses degeneracy of the infinite quantum system with respect to the angular momentum $\hbar m$. For the finite 2d-system one finds a finite degree of degeneracy $\gamma$ for each Landau level in the case of eigenstates (\ref{solution1}) as well. To explain this one has to consider Berry's phase \cite{Ber:84}.

Under QHE conditions, the electric Hall field $\vec{E}=(0,E,0)$, together with the magnetic field $\vec{B}=(0,0,B)$, drives the guiding center of the electron with an average velocity $\vec{<v>}=(v,0,0)$. But the real microscopic charge transfer, which may be illustrated by an adiabatically moved vortex, is rather complicated. It can be characterized by Berry´s phase, which describes here the coupling between rotational and translational movements of the electron.

Starting from the Hamiltonian $H(\vec{r},\vec{R}(t))$, which contains the fast variable $\vec{r}$ coupled with a slow parameter $\vec{R}(t)$ (the guiding center), one uses a wave function for the time evolution
\be
\Psi (t)= \sum\li_n a_n(t) |n(\vec{R}(t))> \label{koeff} \qquad ,
\ee
where $|n(\vec{R}(t))>$ denotes the complete set of eigenfunctions of the so-called snapshot Schr\"odinger equation corresponding to the discrete spectrum $E_n(\vec{R}(t))$. Transitions between the states $|n>$ may be neglected. One then finds for the time dependent coefficients in (\ref{koeff}):
\be
a_n(t) = \exp(-\frac{i}{\hbar}\int\li_0^t E_n(\vec{R}(t'))dt')\; \exp(i\int\li_0^t D_n(\vec{R}(t'))dt') \label{phas}
\ee
where the dynamic and geometric phases are in the two exponential terms of (\ref{phas}). Here
\bea
D_n(\vec{R}(t'))dt' &=& i<n(\vec{R}(t'))|\pderiv{t'}|n(\vec{R}(t'))>dt'\nn\label{form}
&=& i<n(\vec{R})|\nabla_{\vec{R}}|n(\vec{R})>d\vec{R}=\vec{D}_n(\vec{R})d\vec{R} 
\eea
is a 1-form over the $\vec{R}$--parameter space. The matrix element in (\ref{form}) transforms as a vector potential, i.e. for a phase transformation such as
\be
|n> \lra |n'> = e^{i\alpha} |n>
\ee
the equation
\be
\vec{D}_n'=\vec{D}_n - \nabla\alpha\qquad .
\ee
holds. For an arbitrary closed path in the parameter space $\vec{R}(T)=\vec{R}(0)$, the Berry phase
\be 
\theta^n_{\mbox{\scriptsize top}} = \oint \vec{D}_n(\vec{R}) d\vec{R} \label{phas2}
\ee
is, in general, a topological invariant.\footnote{An analogous consideration holds also for unitary transformations.}

Here we notice that a Berry phase unequal to zero is only expected if the parameter space has a nontrivial topological structure, i. e. it must be a multiply connected space. In the language of fiber bundles the fibers are Hilbert spaces on the parameter space which is in our case a 2d-manifold. The 1-form transforms as a connection in a fiber bundle over the parameter space. This corresponds to an Abelian gauge field with $U(1)$-symmetry.\footnote{In case of degeneration of states $|n,k>$ the 1-form is matrix valued $\vec{D}^{kk'}_n(\vec{R})d\vec{R}=i<n,k|\nabla_{\vec{R}}|n,k'>d\vec{R}$ and one has a non-abelian gauge field of the symmetry $U(\delta)$ in a subspace of the dimension $\delta$.}

Identifying $|n(\vec{R})>|_{\vec{R}=0}$ with the states in (\ref{solution1}) one sees that the wave function at $\vec{R}'$ is generated from the wave function at $\vec{R}$ by the following gauge transformation
\be
\Psi_n(\vec{r}-\vec{R})\lra \Psi_n(\vec{r}-\vec{R}') = \exp(-\frac{ie}{\hbar}\int\li_{\vec{R}}^{\vec{R}'} \vec{A}(\vec{R}'')d\vec{R}'')\; \Psi_n(\vec{r}-\vec{R})\; \label{wf1}
\ee
and that the topological phase (\ref{phas2}) is given by
\be
\theta^n_{\mbox{\scriptsize top}} = \frac{e}{\hbar}\oint \vec{A}(\vec{R}) d\vec{R}=\frac{e}{\hbar}\Phi \qquad .
\ee
This expression corresponds to the quantum interference condition for an electron moving around a magnetic flux $\Phi$, or in other words the Aharonov--Bohm effect. Introducing the flux quantum $\Phi_0=\frac{h}{e}$, one finds 
\be
\theta^n_{\mbox{\scriptsize top}} = 2\pi \frac{\Phi}{\Phi_0} = 2\pi \gamma \label{LLentartung}
\ee
which is the equation (\ref{quant1}). This shows that the degree $\gamma$ of degeneracy for Landau levels in the finite plane $L_xL_y$ is equal to the number of flux quanta in this plane, where the filling factor expresses the ratio of the numbers of electrons and fluxes. In such a picture for the IQHE the relevant quasiparticles are composed of an integer number of charges which successively occupy Landau levels and carry together one flux quantum according to $\nu=\frac{N_e}{\gamma}=1,2,\ldots $.

In addition to the orbital motion of electrons under QHE conditions, one has to consider the coupling between the spin and the magnetic field as described by the Pauli Hamiltonian 
\be 
H_P=\left(\frac{1}{2M}\right)(-i\hbar\nabla - e\vec{A})^2 + \left(\frac{e\hbar}{2M}\right)\vec{\beta}\cdot \vec{B} \label{pauli1}
\ee
where $\vec{B}=\nabla\times \vec{A}$ and $\vec{\beta}=(0,0,\pm 1)$ denotes the spin direction. For the 2d-electron there is a representation of (\ref{pauli1}) as square of the Dirac operator \cite{SeSo:86}
\be H_P=\frac{1}{2M} H_D^2 \ee
with
\be H_D=(i\hbar\vec{\sigma}\cdot\nabla+e\vec{\sigma}\cdot\vec{\mbox{\bf A}})=\left(\begin{array}{cc} 0&{/\mkern -10mu \partial}\\{/\mkern -10mu \partial}^+&0 \end{array}\right)\; . \label{dirac1} \ee
and $\vec{\sigma}$ is the vector of Pauli spin matrices. Here we choose the asymmetric gauge $\vec{A}=(-By,0,0)$
and the torus $T^2$ as the underlying manifold. A direct calculation of the spectrum of the operator in (\ref{dirac1}) shows that $\dim\ker {/\mkern -10mu \partial}^+=0$, or in other words, that the ground state is spin--polarized. The result obtained with the Atiyah--Singer index theorem is given by
\be 
\dim\ker H_P=\dim\ker {/\mkern -10mu \partial}=\frac{e}{2\pi\hbar}\int\li_{T^2} B d^2\vec{r}=\frac{\Phi}{\Phi_0}=\gamma \quad . \label{index1}
\ee
This means that the one-particle ground state including spin has the degeneracy $\gamma$ which is equal to the number of flux quanta. According to the well known expression for the one-particle energy
\be
E_{nm\beta}=\hbar\omega_c [ n+\frac{1}{2}(|m|+m)+\frac{1}{2}+\beta\frac{1}{2}] \label{LL1}
\ee
the ground state energy is $E_{0,m,-1}=0$ for $m=-\gamma, \ldots ,0$; $\omega_c=\frac{eB}{M}$.

\setcounter{section}{2}
\section{Many--particle picture of the FQHE}
For the case of FQHE we will consider a time--dependent 2d-configuration $\{\vec{R}_i(t)\}$ of the guiding centers of $N_e$ electrons on a positively charged background. Each electron carries its own Hilbert space represented by the wave functions given in (\ref{solution1}) and completed by the two spin states. Just as in section 2 the underlying manifold for one--particle states is the torus $T^2$.

The many--particle wave functions in the corresponding Fock space have to be antisymmetric with respect to all coordinates $\vec{r}_i$ and all spins $\vec{s}_i$ of the electrons. In principle there are two ways to find these states. The first is to solve the many--particle problem exactly, including all interactions, which is impossible. The second method is to choose the decomposition of the many--particle problem by considering a product of one--particle wave functions and encoding the interactions in the geometric structure of the underlying manifold. 

The first method came about when Laughlin approached the many-particle state intuitively. He wrote down a trial wave function to produce filling factors $\nu=\frac{1}{m}$, $m=1,3,5\ldots$ \cite{Lau:83}:
\be
\chi_{\frac{1}{m}}=\prod\li_{\stackrel{j,k}{(j<k)}} (z_j-z_k)^{m-1}\chi_1 \label{lau1}
\ee
with
\be
\chi_1=\prod\li_{\stackrel{j,k}{(j<k)}} (z_j-z_k)\exp(-\frac{1}{4}\sum\li_{i=1}^{N_e} |z_i|^2) \label{lau2} \qquad .
\ee
The many particle state in (\ref{lau1}) contains $m$ fluxes per electron. The electron positions are denoted by dimensionless (in units of $l_0$) complex variables $z_j=x_j+iy_j$.
This wave function is antisymmetric in the coordinate part but symmetric with respect to the spin i.e. all electrons are spin-polarized. Later, Jain showed that the construction of other filling factors $\nu=\frac{p}{q}$ is possible if higher Landau levels are included in the correlated electron state \cite{Jai:89}
\be
\chi_{\nu}=\prod\li_{\stackrel{j,k}{(j<k)}} (z_j-z_k)^{m-1}\chi_p \label{jain1}
\qquad . \ee 
Here $\chi_p$, $p=1,2,3,\ldots $ is for $p>1$ an unknown antisymmetric function describing the $p$-th IQHE state. For the $p$-th Landau level $\frac{1}{p}$ fluxes per electron are realized. In generalizing from (\ref{lau1}) to (\ref{jain1}), one has to add the even number of $(m-1)$ fluxes to find $\frac{1}{\nu}=\frac{1}{p}+(m-1)$. The fractional filling factor is then found to be\footnote{In order to realize all experimentally observed fractions one has to permit $p=\pm1,\pm 2,\ldots $ leading to $\nu=\frac{p}{(m-1)p\pm 1}$.}
\be
\nu = \frac{p}{(m-1)p+1} \qquad .
\ee
Indeed for semiconductors, with an effective electron mass $M^*< M$ and a dielectric constant $\epsilon>1$, the comparison of the distance $\hbar\omega_c$ between Landau levels with the characteristic Coulomb energy $\frac{e^2}{4\pi\epsilon_0\epsilon l_0}$, for realistic magnetic fields $B\sim 10T$, indicates that Landau level mixing cannot in general be neglected for understanding the many-particle state. Furthermore, if we consider the one--particle problem including spin, it follows from (\ref{LL1}) that a change in spin leads to higher Landau levels, so that the ground state energy with spin $\frac{1}{2}$ is equal to the energy of the first Landau level with spin $-\frac{1}{2}$. 

In the following we will employ the second method to acquire a topological many--particle description of the FQHE states.
We start with the one--particle picture given in the previous section. The electron configuration $\{ R_i(t)\}$ forms the underlying manifold. The geometry and topology of this manifold are changed with respect to the interaction of particles. This change restricts the number of possible wave functions. Only wave functions with specific zeros or nodes are allowed. The mathematical way to express this idea is given by the theory of fiber bundles. 

We consider a complex line bundle $L$ over $T^2$ with trivialisation $(U_i,\phi_i)$. The set of sections $\sect{U_i,L}$ over the chart $U_i$ forms a Hilbert space. Each one--particle wave function $\phi_i$ is an element of $\sect{U_i,L}$. That means that we are choosing a trivialisation with only one particle in each chart $U_i$ and have
\be
T^2 = \bigcup\li_{i=1}^{N_e} U_i\qquad\mbox{with}\quad U_i\cap U_j\not= \emptyset \quad \forall i,j \qquad .
\ee
The interaction is encoded in the geometrical and topological properties of the transition function $g_{ij}:U_i\cap U_j\times\C\rightarrow U_i\cap U_j\times\C$. One point in each chart describes the guiding center of the particle. In the case of the torus $T^2$ one needs a minimum of 4 charts to represent this manifold. Our construction holds for an arbitrary particle number $N_e\geq 4$. The unsolved problem consists in transporting a wave function from one chart to another one. The transport of the wave function is described by the connection in the bundle. To determine this connection we construct the principal fiber bundle $P(L)$ associated with the bundle $L$. 

There is an ambiguity in the local definition of the wave function. The physical interpretation of the wave function implies that it is invariant with respect to global phase transformations. All phase transformations form a group $U(1)$. Therefore, we introduce a local phase change. A local phase change implies that there exists a relative phase between two particles. This local phase gives rise to the existence of a gauge field, which is, in the mathematical sense equivalent to the connection of the $U(1)$-principal fiber bundle $P(L)$ associated with $L$. Two principal fiber bundles over different spaces are equivalent (isomorphic) if there exists a continuous deformation (homotopy) from the one space to the other. Such a deformation does not change the overlap between two charts. This is the method underlying the classification of fiber bundles. Fortunately, the classification is complete and the corresponding invariants are given as polynomials of the curvature. 

In implementing the method described above, let $\omega$ be a connection in $P(L)$. The connections $\omega|_{U_i}$ and $\omega|_{U_j}$ are restrictions of $\omega$. The relation between both restrictions of $\omega$ is given by the gauge transformation
\be
\omega|_{U_j}= \omega|_{U_i} + g_{ij}^{-1} dg_{ij}
\ee
where $g_{ij}:U_i\cap U_j\rightarrow U(1)=S^1$ is the transition function in $P(L)$.
All $U(1)$--principal fiber bundles over $T^2$ will be classified by the first Chern class $c_1$ as an element of the second cohomology group $H^2(T^2,\Z)=\Z$ \cite{Ste:51,MSt:74}. The expression of $c_1$ is given by \cite{EgGiHa:80}
\bea
c_1=\frac{i}{2\pi}d\omega = \frac{i}{2\pi}\Omega \label{Chern1}\qquad , \\
\frac{i}{2\pi}\int\li_{T^2}\Omega=C_1 \in\Z \label{Chern2}
\eea
where $\Omega=d\omega$ is the curvature and $C_1$ is the Chern number. Now we choose two regions $G_1,G_2$ of $T^2$, where $T^2=G_1\cup G_2$ and $\omega|_{G_1}=0$ \footnote{For our consideration here the remaining two regions of the complete decomposition of the torus are not needed.}. The connection on $G_2$ is given by
\be
\omega|_{G_2}=\omega|_{G_1} + g_{12}^{-1}dg_{12}= g_{12}^{-1}dg_{12} \label{gaugetrafo}
\ee
with the transition function $g_{12}:G_1\cap G_2\lra U(1)$. Next we consider a curve $\zeta$, where $[\zeta]\in H_1(T^2,\Z)$, as shown in the figure 1.\\[0.4cm]
\begin{figure}
\hspace*{4cm}\begin{minipage}{8cm}
\psfig{figure=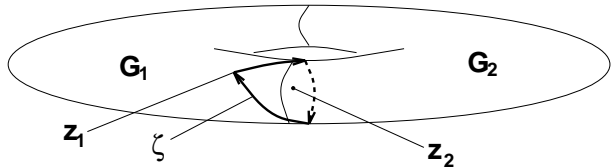,width=8cm}
\end{minipage}
\caption{Noncontractable curve $\zeta$ on the torus $T^2$ for two interacting particles.}
\end{figure}\\[0.4cm]
Using standard methods of cohomology theory one obtains for the integral of the expression (\ref{gaugetrafo})
\be
\frac{i}{2\pi} \int\li_\zeta g_{12}^{-1}dg_{12} \in \Z
\ee
which is related to the Chern number. If we choose two particles at positions $Z_1,Z_2\in\C$ in $G_1$ and $G_2$, respectively, (see picture) then it follows from the residuum theorem and the discussion given in \cite{AlEe:57}, that
\be
\frac{i}{2\pi} \int\li_\zeta g_{12}^{-1}dg_{12}=\frac{i}{2\pi}\int\li_{|Z_1-Z_2|=1} \frac{dZ_1}{Z_2-Z_1} = 1 \qquad .
\ee
The uniqueness of this relation follows from the uniqueness of the residuum theorem. 

The interaction term for two particles is thus determined by
\be
\omega|_{G_2}= \frac{dZ_1}{Z_2-Z_1} \label{interact1}
\ee
which leads to a shift in the complex vector potential for the $i$-th particle interacting with all other particles
\be
A(Z_i)\lra A(Z_i) + \frac{\hbar}{e}\sum\li_{j\atop i\not= j} \frac{1}{Z_j-Z_i} \label{pot1} \qquad .
\ee
This potential is similar to but different from the Coulomb potential \cite{AroSchWil:84.2}. Namely, it is a complex number with nontrivial phase, and the interaction appears now in covariant derivatives \cite{FlVa:94}. Furthermore, it does  not contribute to the curvature or the magnetic field, except for $Z_i=Z_j$.

Obviously, the wave function as a section of $L$ was assumed to be non-degenerate. To generalize the situation to degenerate states we have to use a complex vector bundle $V$ of rank $\delta$ over $T^2$ with $\delta$ as the degree of degeneracy. The set of sections $\sect{T^2,V}$ are the wave functions of the degenerate problem. This bundle is associated with a $U(\delta)$-principal fiber bundle $P(V)$ over $T^2$. The connection $\theta$ in $P(V)$ is matrix--valued, where the matrix are elements of the Lie algebra of $U(\delta)$. Fortunately, the classification does not change in principle, and we obtain 
\be
\frac{i}{2\pi}\int\li_{T^2} \tr\Theta =C_1 \in\Z 
\ee
where $\Theta=d\theta+\theta\wedge\theta$ is the curvature. Taking the trace of $\Theta$ reduces the problem from one involving the whole $U(\delta)$-group to one involving only the $U(1)$-subgroup. So we obtain for the degenerate case the same interaction term (\ref{interact1}) as in the non--degenerate case. 

Additionally, we want to remember the relationship between fiber bundle theory and the topological phase considered in section 2. In \cite{BoBoKe:92} it is shown that the Berry phase is generated by the connection of a universal bundle. But there is a local diffeomorphism which relates the Berry connection to the connection discussed above. So our construction makes sense and the interaction can be interpreted in the picture of topological phases.

\section{Consideration of two spin structures}
In the previous section the substantial interaction term (\ref{interact1}) has been found by pure topological arguments. Now we will extend our consideration on the line of classification of fiber bundles to the spin. The nontrivial topology of the relevant fiber bundles caused by the characteristic of the underlying manifold, is also responsible for the existence of different types of fermions according to the spinor form of the wave functions.

These so--called ``twisted spinors'' were  first investigated
in quantum field theory   and in particular in quantum
gravity \cite{Is1:78,Is2:78}. In 1979 Petry introduced the notion of ``exotic spinors'' in the context of superconductivity \cite{Pet:79}.

In fact, there exist as many inequivalent spinors as there are elements in the
cohomology group $ H^1(M,{\Z}_2) $, where $M \,$ is the configuration space. This particular
group may be used for representing the inequivalent nontrivial
circles (cocycles) and all the allowed combinations of them. Therefore there is an additional degree of freedom for fermionic particles on our multiply connected space. Intuitively, one might visualize this new degree of freedom as something similar to a M\"obius band, i.e. spinors may be twisted around the nontrivial circles. Classically one is inclined to think of these different spinors as describing different
particles. In the quantum picture there
should be a sum over all possibilities and a new partition function \cite{AvIs:79}.

Under some additional assumptions on the topology
 of $M \,$, it is possible to translate the different spinors into one chosen spinor frame \cite{Pet:79}.
Essentially, the twisted spinors behave like 
normal spinors after translation, the only important difference being that the covariant 
derivative acting on these spinors has to be changed by an additional 1-form 
which makes exotic spinors very interesting in the context of FQHE. 

The spin group in two dimensions is an Abelian group $SO(2)=U(1)$. Spinors 
can be characterized by the way they behave under
rotations. If one needs two full rotations to
transform a particle back to the starting point, then this particle is
a fermion with spin $\frac{1}{2}$.
In more than two dimensions this property is connected with the
universal covering $Spin(n)$ of $SO(n)$, for $n>2$. Because 
$\pi_1(SO(n))=\Z_2$, we obtain a 2-to-1 mapping between $Spin(n)$ and 
$SO(n)$.\footnote{A 2-to-1 map is a map with kernel $\Z_2$.}
In the following we construct such
a 2-to-1 mapping for the two--dimensional situation 
\be 
\map{f}{\C}{\C}{z}{z^2}\qquad . \label{ueberlagerung}
\ee
Furthermore, we consider the tangential bundle $TM$ of
a Riemannian manifold $M$ and try to
construct the bundle splitting $TM=S\otimes S$.\footnote{The Riemannian manifold is a
  two--dimensional, connected, open manifold.} Such splitting exists
if the second Stiefel--Whitney class $w_2(M)\in H^2(M,\Z_2)$ vanishes,
which is always the case in two dimensions. The splitting can be compared
to the quadratic map $f$ described by (\ref{ueberlagerung}). 
Sections of the complex
line bundle $S$ are spinors. Now we ask: How many splittings of the type $TM=S\otimes S$ do exist? The
answer is : the existence is given by the $H^2(M,\Z_2)$
and the dimension of $H^1(M,\Z_2)$ determines the number of
splittings \cite{Hi:74}. Every splitting corresponds to another spin structure, which means that 
the $H^1(M,\Z_2)$-group measures the different spin
structures. 

In section 3 we constructed the interaction term (\ref{interact1}). 
This term produces a singularity in the sense that every
particle sees the other particle as a hole and two particles produce two 
holes\cite{AlEe:57}. Therefore, the topology of
$T^2$ changes to $T^2\backslash \{Z_1,Z_2\}$ with $Z_1,Z_2$ as removed
points.  This produces two generators in $H_1(M,\Z)$. If we count all 
possibilities, four different cases are obtained.  

Now we will investigate the relationship between
two spinors in different spin structures. Fortunately, this relationship 
is calculated in \cite{Pet:79} for $3+1$ dimensions, which can be generalized
to all other dimensions\footnote{There is only one restriction: the
  $H_1(M,\Z)$ does not have any 2--torsion.}. The covariant derivative
$\nabla_Y^{e} \,$ of exotic spinors $\psi^{e} \,$ in the direction of a
vector field $Y \,$ is given by
\begin{eqnarray}\label{uff}
\nabla^{e}_{Y}   \psi^{e}  = \nabla_{Y}  \psi^{e} 
 -  \frac{1}{2} \,   \big[ \iota(Y) \,   \, \lambda^{-1} \, \mbox{\rm d} \,\lambda \, \,\big]   \,\psi^{e}
 \end{eqnarray}
where $\nabla_Y \,$ is the usual spin connection and $\iota \,$ denotes
the contraction of a 1-form with the given vector field. Note the
factor $ \frac{1}{2} \, $ in formula (\ref{uff}), which is a consequence of the
mathematics. In (\ref{uff}), the 1-form $\lambda^{-1} \, \mbox{d} \,\lambda \,$ 
with $\lambda :M \, \longrightarrow \, U(1) \,$ is closed but not exact.
It is chosen in such a way that it
represents the particular nontrivial element in the above mentioned
cohomology group responsible for its existence. Therefore the
integral over circles of the 1-form $\frac{i}{2 \pi} \, \lambda^{-1} \, \mbox{d} \,\lambda \,$ in the configuration space 
takes only values $ 1 \,$ or $ 0 \, $ depending on whether the circle
can be deformed to the particular non--contractible circle or not.
Let $\zeta$ be a curve surrounding a hole induced by the interaction term, i.e. 
$[\zeta]\in H_1(M,\Z)$. Then it follows that
\be
\frac{i}{2 \pi} \, \int\li_{\zeta}  \lambda^{-1} \, \mbox{d} \,\lambda =1 \in\Z_2 \qquad .
\ee
Considering the two--particle interaction corresponding to (\ref{interact1}), 
we have to remove two points from the configuration space. 
However, for further topological considerations it is better to remove a little disk instead of a point to achieve a compact configuration space $M$ again.
Therefore, two new generators of $H^1(M,{Z\!\!\!Z}_2)$ appear in correspondence to the number of removed disks. Since the translation procedure follows the group law of $H^1(M,{Z\!\!\!Z}_2) \,$, 
we get ``classically'' three types of exotic spinors with additional 
1-forms:  $\frac{1}{2} \, \lambda^{-1} \, \mbox{\rm d} \,\lambda \,$ , 
$\frac{1}{2} \, \alpha^{-1} \, \mbox{\rm d} \,\alpha \,$ and 
 $\frac{1}{2} \, \left( \lambda^{-1} \, \mbox{\rm d} \,\lambda \, 
+ \,  \alpha^{-1} \, \mbox{\rm d} \,\alpha \right) \,$. 
Due to (\ref{uff}), the derivative in in the Hamiltonian (\ref{pauli1}) is shifted 
for the different exotic spinors by the corresponding 1-form given above. 
Following  \cite{Pet:79} and \cite{Is1:78} we interprete this 1-form as a global vector 
potential generated by the nontrivial topology, which does not contribute to the internal 
magnetic field (because the 1-form is always closed ).  We stress that Stokes theorem
does not apply here in the usual sense due to the nontrivial topology. 
Therefore, we define the total flux $\Phi$ through our system
as the line integral  of the relevant connection 
along the boundary of the configuration space. 
\begin{eqnarray}
\Phi = \oint\limits_{\partial M} A(z) dz = \oint\limits_{\partial D_1\cup \partial D_2} A(z)dz = \frac{h}{e}\cdot q \qquad q=1,2,\ldots
\end{eqnarray}
where $\partial D_1,\partial D_2$ denote the boundaries of the removed disks. If we now assume that all different  spinors (spin structures) contribute equally, then the following equations hold:
\begin{eqnarray}
\Phi\cdot\frac{e}{h} &=& \oint\limits_{\partial D_1\cup \partial D_2} \frac{e}{h}A(z)dz + \oint\limits_{\partial D_1\cup \partial D_2} (\frac{e}{h}A(z)dz - \frac{i}{4\pi}\lambda^{-1}\mbox{\rm d}\lambda -
\frac{i}{4\pi}\alpha^{-1}\mbox{\rm d}\alpha) \nonumber \\ &+& \oint\limits_{\partial D_1} (\frac{e}{h}A(z)dz - \frac{i}{4\pi}\lambda^{-1}\mbox{\rm d}\lambda)+\oint\limits_{\partial D_2} (\frac{e}{h}A(z)dz - \frac{i}{4\pi}\alpha^{-1}\mbox{\rm d}\alpha)\nonumber \\
&=& q + (q - \frac{1}{2}-\frac{1}{2}) + (q - \frac{1}{2}) + (q - \frac{1}{2})=4q-2
\end{eqnarray}
where $\Phi\cdot\frac{e}{h}$ is the number of flux quanta. 

Since there are two particles involved, the denominator of the filling factor $\nu$ is given by
\begin{eqnarray}
\frac{1}{2}\Phi\cdot\frac{e}{h} = 2q-1 \sim \frac{1}{\nu} \qquad .
\end{eqnarray}
According to \cite{Ver:91}, it seems natural to assume that interaction is dominated by forming 
pairs; then the total flux produced  is just the flux 
$ N (4q - 2 ) \,$ for $ N=\frac{1}{2}N_e$ pairs of fermions leading again to the factor 
$2q - 1 \,$ in $ \frac{1}{\nu} \,$.\\
The discussed model of the topological origin of some aspects of electron-electron interaction should complete the usual composite fermion picture. 
The model also provides a possible explanation
 of the odd denominator in the filling factor.

\section{Discussion of the Hall conductivity}
In this section we want to investigate the Hall conductivity for the general case of an interacting electron system with a degenerate ground state. 
The corresponding expression will be a topological invariant. First, we reproduce the ``classical'' result, i.e. the Hall conductivity as the 
first Chern number of an $U(1)$-principal fiber bundle over 
$T^2$ \cite{NiuThoWu:85,AvrSei:85}. 

Let us consider the Hamiltonian $\hat{h}$ together with the eigenvalues $E_n$, which satisfy the stationary Schr\"odinger equation
\be
\hat{h}|n> =E_n|n>\qquad , \label{eigenwertgleichung1}
\ee
where $|n>=|n(\vec{R})>$ are parameter--dependent states of a two--dimensional parameter space $P$. The topological phase is generated by the Berry connection
\be
\omega_{mn}= <m|d|n>=<m|\partial_i|n> dR^i \equiv <m|\pderiv{R^i}|n>dR^i
\ee
where $i=1,2$. Normalization $<m|n>=\delta_{mn}$ yields
\be
\omega_{mn}= - \omega^*_{mn}
\ee
and the curvature is given by
\be
\Omega_{mn}=d\omega_{mn}=\sum\li_k \omega_{mk}^* \wedge \omega_{kn}= - \sum\li_k \omega_{mk} \wedge \omega_{kn} \label{curv1} \qquad .
\ee
Formula (\ref{curv1}) may be expressed in terms of the derivatives of the Hamiltonian $\hat{h}$. Using
\be
<m|\partial_i(\hat{h}|n>)=<m|\partial_i \hat{h}|n> + <m|\hat{h}\partial_i|n>
\ee
and (\ref{eigenwertgleichung1}), one obtains 
\bea
<m|\partial_i|n>&=& \frac{<m|\partial_i \hat{h}|n>}{E_m-E_n} \\
<n|\partial_i|n>&=& 0 \qquad .
\eea
The diagonal part of the curvature is given by
\be 
\Omega_{nn}=\sum\li_{k\not= n}\frac{<n|\partial_1\hat{h}|k><k|\partial_2\hat{h}|n> - <n|\partial_2\hat{h}|k><k|\partial_1\hat{h}|n>}{(E_n-E_k)^2} dR^1\wedge dR^2 \label{curv2} \qquad .
\ee
The integration of expression (\ref{curv2}) gives an integer result according to formula (\ref{Chern2}). 

Within the concept of adiabatic transport the many--particle Hamiltonian for the current carrying QHE-system is
\be 
H=\frac{1}{2M}\sum\li_{i=1}^{N_e} (\vec{p}_i+\hbar \vec{\Phi})^2 +\sum\li_{i<j}^{N_e} V_c(\vec{r}_i-\vec{r}_j)+ \sum\li_i^{N_e} V_b(\vec{r}_i) 
\ee
where $\vec{p}_i$ is the canonical momentum. Here $V_c$ is the Coulomb potential and $V_b$ describes the random background potential, which we will ignore in the following. The macroscopic flux variable $\vec{\Phi}=(\Phi_x/L_x,\Phi_y/L_y)$ on the torus is used instead of a statistical time--dependent vector potential $\vec{A}(\vec{r}_i,t)$. The vector $\vec{\Phi}$ is connected with stationary currents flowing through the system, ($\Phi_x,\Phi_y)$ are dimensionless in units of $\hbar/e$.

The Kubo formula for calculating the non--diagonal part of the conductivity is given by
\be 
\sigma_{xy}=\frac{ie^2 \hbar}{L_xL_y}\sum\li_n \frac{<0|v_x|n><n|v_y|0> -<0|v_y|n><n|v_x|0>}{(E_0-E_n)^2} 
\ee
where $\vec{v}=(v_x,v_y)$ is the velocity operator of the center of mass of the electron system. Here $|0>$ and $|n>$ denote the many-particle ground and excited states, respectively. Using the relation
\be
\vec{v}=\frac{1}{\hbar}\frac{\partial H}{\partial\vec{\Phi}} 
\ee
it follows that
\be 
\sigma_{xy}=\frac{ie^2}{\hbar} \sum\li_n \frac{<0|\frac{\partial H}{\partial \Phi_x}|n><n|\frac{\partial H}{\partial \Phi_y}|0> - <0|\frac{\partial H}{\partial \Phi_y}|n><n|\frac{\partial H}{\partial \Phi_x}|0>}{(E_0-E_n)^2} \qquad .
\ee
Comparing this expression with (\ref{curv2}), one obtains
\be 
\sigma_{xy}\frac{d\Phi_x}{2\pi} \frac{d\Phi_y}{2\pi}=\frac{e^2}{h}\frac{i}{2\pi}\Omega_{00}=\frac{e^2}{h} c_1  \label{conduct1}
\ee
where $c_1$ is the Chern class (\ref{Chern1}) and is proportional to the local conductivity. Integrating (\ref{conduct1}) yields an expression for the macroscopic Hall conductivity
\be 
<\sigma_{xy}>=\frac{e^2}{h}\int\li_{T^2} c_1=\frac{e^2}{h} p \qquad \mbox{with} \quad p\in \Z \label{conduct2}\qquad ,
\ee
which is quantized. The calculation given here is only valid for a large electron number $N_e$. It reflects the many--electron picture generally in the states $|0>,|n>$, and indirectly in the flux variable $\vec{\Phi}$, which is always an averaged quantity. For $q$-fold degenerate ground states $|0>$ formula (\ref{conduct2}) must be completed by the factor $\frac{1}{q}$ producing the filling factor $\nu=\frac{p}{q}$.

To investigate the many-particle problem, a quantum field theory calculation may be prefered. The conductivity is proportional to the current--current correlator, which can be expressed in terms of variational derivatives of the generating functional. To do this, we describe the quantum Hall states as solutions of the Dirac equation
\be 
\dirac_\theta \psi=i\hbar\gamma^\mu(\partial_\mu + \theta_\mu)\psi=0 
\ee
where $\theta_\mu$ is the connection of the $U(\delta)$-principal fiber bundle, and includes both the vector potential of the magnetic field and the particle--particle interaction term. Next we consider the general situation of a closed connected Riemannian manifold $M$. The generating functional parametrized by the connection $\theta$, is given by
\be 
{\cal Z}(\theta)=\int D\psi D\overline{\psi} \exp (-\frac{1}{\hbar}S[\psi,\overline{\psi},\theta])\label{action}
\ee
with the action
\be S[\psi,\overline{\psi},\theta]=\int\li_M d^2 x \overline{\psi}(x)\dirac_\theta\psi(x) \qquad . \ee
The current--current correlator can be calculated from
\be
\frac{1}{{\cal Z}}\frac{\delta^2{\cal Z}}{\delta \theta^\mu\delta \theta^\nu}=\frac{\int D\psi D\overline{\psi} j_\mu j_\nu \exp (-S)}{\int D\psi D\overline{\psi} \exp (-S)}\equiv <j_\mu j_\nu> \label{functcorr}
\ee
where the 1d-Dirac current is given by
\be
j^\mu_{ab}=\overline{\psi}\gamma^\mu(1+\gamma^5)\psi t_{ab}\qquad \mu=1,2 \qquad .
\ee
Here, $t_{ab}$ is the generator of the Lie algebra of the $U(\delta)$ group. These formulas allow for electron--electron interaction in the fixed gauge field $\theta$. The correlation between the two currents is obviously dependent on this gauge field. The expression in (\ref{action}) can be calculated by using standard methods which leads to
\be
{\cal Z}(\theta)=\frac{1}{\hbar}(\det (\dirac^*_\theta \dirac_\theta))^{\frac{1}{2}} \qquad .
\ee
Using the QHE condition 
\be
<j_1 j_2>\equiv <j_xj_y>=-<j_yj_x>
\ee
the correlator in (\ref{functcorr}) is
\be 
<j_x(\vec{z}) j_y(\vec{z}^\prime)>=\hbar^2\, \tr \left(\frac{\gamma_5}{\dirac_\theta(\vec{z})\dirac_\theta(\vec{z}^\prime)}\right)\delta(\vec{z}-\vec{z}_0)\delta(\vec{z}^{\, \prime} - \vec{z}_0)
\label{corr1}
\ee
where $\vec{z}=(ct,x)$ and $\vec{z}^{\,\prime} (ct^\prime,y)$ are the coordinates on a torus $T^2$. 
The integration over $\vec{z},\vec{z}^\prime$ leads to the macroscopic formula
\be
<<j_x j_y>>=\int\li_{T^2} d^2z d^2z^\prime <j_x(\vec{z}) j_y(\vec{z}^\prime)> = \hbar^2\, \tr \left(\frac{\gamma_5}{\dirac^2_\theta(\vec{z}_0)}\right)\qquad .
\ee
Because the correlator is divergent, one has to regularize the expression in (\ref{corr1}) by 
setting
\be 
\tr\left(\frac{\gamma_5}{\dirac^2_\theta}\right) = \lim_{t\lra 0} \tr\left(\frac{\gamma_5}{t^2+\dirac^2_\theta} \right) \qquad .
\ee
Employing the heat kernel expansion \cite{bvg:90} the following expression is obtained:
\bea
<<j_x j_y>> &=& c^2 L_x L_y \int\li_{T^2} \frac{dt}{2\pi}\frac{dt^\prime}{2\pi} <j_x(t) j_y(t^\prime)> \\
&=& \frac{L_xL_y}{4\pi^2}\int\li_{T^2} \tr\Theta \label{functcorr2} 
\eea
where $\Theta=d\theta +\theta\wedge\theta$ is the curvature. 
The remarkable fact is that the calculation of the correlator by use of the heat kernel expansion leads to a topological invariant known as Chern class. The mathematical idea consists in using the square of the Dirac operator instead of the Laplace operator. There is a correspondence between the formal solution of the heat equation and the Chern character with respect to a suitable superconnection given by the Dirac operator. A comparison with the classical Kubo formula in the beginning of this section leads to the identity
\be
<\sigma_{xy}>= \frac{ie^2}{\hbar\, L_xL_y} <<j_x j_y>> \qquad .
\ee
The calculation is done with respect to only one of the $\delta$ degenerate states. Averaging over all states yields the additional factor $\frac{1}{\delta}$ in (\ref{functcorr2}).
Thus, we obtain for the fractional Hall conductivity: 
\be 
<\sigma_{xy}> = \frac{e^2}{h} \frac{w}{\delta} \qquad w,\delta\in \Z \label{leitf3} 
\ee
with
\[ w = \frac{i}{2\pi}\int\li_{T^2} \tr\Theta \]
whereas the real part of the longitudinal conductivity vanishes. In (\ref{leitf3}) $w$ is known as the winding number of the self-consistent gauge field, whereas $\delta$ denote the winding number of the magnetic field. The conductivity fraction is given by the ratio of two winding numbers, and is a topological invariant.

The calculation can be reduced to the case of a $U(1)$-subgroup. Because of the diffeomorphism $U(\delta)\simeq U(1)\times SU(\delta)$, only the $U(1)$-group contributes to the invariant\footnote{The $SU(\delta)$-group contains all traceless matrices and will be annihilated by the trace.}. This means that
\be
<\sigma_{xy}> =\frac{e^2}{h}\; \frac{1}{\delta}\; \frac{i}{2\pi}\int\li_{T^2} d\omega
\ee
where $\omega$ is the appropriate $U(1)$-connection representing the $U(1)$-subgroup.

At the end of this section we want to discuss the ($2+1$)-dimensional interpretation of the Hall conductivity. Starting from the time-dependent Pauli equation
\be
i\hbar \pderiv{t} \psi = H_P \psi = \frac{1}{2M}H_D^2 \psi 
\ee
the conductivity $<\sigma_{xy}>$ is now given by 
\be
\frac{h}{e^2}<\sigma_{xy}> = \frac{i}{2\pi}\int\li_G d\omega = \frac{i}{2\pi}\oint\li_{\zeta=\partial G} \omega \label{coduct2}
\ee
where $G\subset T^2$ is a suitable subset of the torus with non-vanishing connection and $\zeta$ is the curve given by the picture in section 3. The conductivity, being a topological invariant, does not change. Therefore an extension of the formula (\ref{conduct2}) obtained from the adiabatic curvature is possible. The configuration space is now a ($2+1$)-dimensional manifold. We choose the Chern-Simons action over ${\cal F}=T^2\times S^1$ as a simple gauge invariant where $S^1$ is the periodic time. The extension of the connection $\omega$ is denoted by the same symbol, and the action is:
\be
S[\omega]=\hbar\int\li_{{\cal F}} \omega\wedge d\omega \qquad .
\ee
Now we consider the expectation value
\bea
<\exp\left(i\frac{h}{e^2}\sigma_{xy}\right)> = \prod\li_{a=1}^N<\exp(\frac{i}{2\pi}\oint\li_{\zeta_a} \omega)>=\int D\omega \exp\left(\frac{i}{\hbar}S[\omega]\right) \prod\li_{a=1}^N\exp(\frac{i}{2\pi}\oint\li_{\zeta_a} \omega)
\eea
where $a$ denotes all curves of the particles. The value of this average is known \cite{Pol:88}, yielding the nice result
\be
<\sigma_{xy}>=\frac{e^2}{h} l
\ee
where $l$ is the linking number of the curves $\zeta_a$. This linking may be interpreted as the interaction between the particles. The number $l$ is also a fraction, in agreement with the calculations given above. 

\section{Summary and Conclusions}
In this paper we presented a topological discussion of the quantized Hall conductivity. Starting from the fact that the stationary one-particle problem for a 2d-electron in a perpendicular magnetic field has strong solutions, the many-particle correlation including the spin is treated by use of Berry's connection. The mathematical basis for this concept is given in the theory of fiber bundles. The underlying manifold is considered to be the configuration space and the time. In adiabatic transport theory the latter is eliminated but in general it may be considered in the sense of a $(2+1)$-dimensional field theory. 

The main idea of our investigation consists in encoding the relevant physical interaction in characteristics of the underlying manifold. Using one chart per electron considered to be clothed by the magnetic field, particle-particle interaction is implied in characteristic transition functions. To investigate the local mathematical properties of the transition functions characterized by zeros or nodes, Morse theory should be employed. 

An expected quasi-long range order of the macroscopic quantum state seems to be reflected by the topology of flux quantization which enables phase transitions between quantum liquid and quantum crystal states.

The quantized Hall conductivity known to be a topological invariant may be characterized by the linking number being in general a rational fraction of two winding numbers. So our theory confirms the quasi-particle picture of FQHE where an even or odd number of flux quanta has been attached to fermions or bosons, respectively. 

Recognizing that composite particles behave like dressed charges in a weak magnetic field, one should expect a direct mapping between FQHE and IQHE as a problem of further investigation.

An alternative approach to FQHE theory by edge excitation (see M. Stone in \cite{Sto:92}) is recently shown to be compatible to the bulk picture due to ${\cal W}_{1+\infty}$ as the underlying symmetry \cite{FlVa:94}.

\end{document}